# The Dangerous Dogmas of Software Engineering

Paul Ralph
Lancaster University
Lancaster, LA1 4YX
United Kingdom
+44 1524 594379
paul@paulralph.name

Briony J Oates
Teesside University
Middlesbrough, TS1 3BA
United Kingdom
+44 1642 342663
B.J.Oates@tees.ac.uk

## ABSTRACT
To legitimize itself as a scientific discipline, the software engineering academic community must let go of its non-empirical dogmas. A dogma is belief held regardless of evidence. This paper analyzes the nature and detrimental effects of four software engineering dogmas – 1) the belief that software has "requirements"; 2) the division of software engineering tasks into analysis, design, coding and testing; 3) the belief that software engineering is predominantly concerned with designing "software" systems; 4) the belief that software engineering follows methods effectively. Deconstructing these dogmas reveals that they each oversimplify and over-rationalize aspects of software engineering practice, which obscures underlying phenomena and misleads researchers and practitioners. Evidenced-based practice is analyzed as a means to expose and repudiate non-empirical dogmas. This analysis results in several novel recommendations for overcoming the practical challenges of evidence-based practice.

## Categories and Subject Descriptors
D.2.0, D.2.1, D.2.10 [**Software Engineering**]: General, Requirements Specifications, Design Methodologies.

## 1. INTRODUCTION

*If you can't make engineering decisions based on data, then make engineering decisions that result in data.*
-Kent Beck

Bats are not blind. Sunflowers do not turn with the sun. Shaving does not make hair grow back thicker. Bulls are not enraged by the color red. Humans are not limited to five senses. Stomach ulcers are not caused by stress. Lemmings do not mindlessly follow each other off cliffs. Deoxygenated blood is not blue. Water does not drain in opposite directions in the northern and southern hemispheres. And Thomas Edison did not invent the lightbulb. The prevalence of these and other common misconceptions [1] leads us to wonder what misconceptions might concern software engineering (SE).

Of course, not all misconceptions are equally important. Some misconceptions are easily rectified, e.g., most people confronted with the idea that our sense of equilibrioception (balance) does not classify neatly into the traditional five senses willingly revise their belief about us having just five senses. In contrast, some beliefs persist regardless of evidence, e.g., faced with extensive evidence of climate change many people continue to believe it is an elaborate hoax. When a belief is held regardless of evidence, it becomes a *dogma*. Moreover, some dogmas and misconceptions are basically harmless, while others, e.g., the belief that vaccines cause autism, can be fatal. We can therefore narrow our scope and propose the following research question:

**Research Question:** *What dangerous dogmas pervade the software engineering discourse?*

Here, a *dangerous dogma* is a belief that harms researchers or practitioners in some way and resists empirical challenge. The ICSE 2014 call for submissions (http://2014.icse-conferences.org/research) explicitly invites "papers that assess the state of the art in the field, its research trajectory, and *core assumptions that may or may not hold* in the future" (italics added). This paper consequently examines several core assumptions (dogmas) in software engineering (SE) that endure despite being logically problematic and contrary to many empirical observations.

Exploring these dogmas is important as they fundamentally undermine SE practice and research. The SE discipline can only progress if practitioners and researchers are willing to recognise dangerous dogmas and modify their beliefs according to the empirical evidence.

We therefore proceed by identifying and deconstructing some of SE's dangerous dogmas (§2). We then explain how SE might transcend dogmas by moving toward evidenced-based practice (§3). Section 4 concludes the paper by summarizing its contributions and limitations.

## 2. UNEMPIRICAL DOGMAS OF SOFTWARE ENGINEERING

In this section we discuss four dangerous dogmas in SE. In each case we identify a commonly-held belief, demonstrate that it is a dogma by showing how it is both logically problematic and contradicted by empirical evidence, discuss the continuing appeal of the dogma and then explain how it is harmful.

### 2.1. Software Systems are Designed to Satisfy Requirements

It is widely believed that software systems are designed to satisfy requirements, where a requirement is a feature or property of a system that is necessary for success [2], [3]. Belief in software requirements is evident in educational curricula [4]-[6], textbooks, industry standards [7], [8], bodies of knowledge (e.g. [3]), software development methods [9]-[11] and the extensive literature on requirements engineering. The idea of specifying requirements independently of designing and building a system is solidified by management pressure to outsource, conceptualize development as independent phases (§2.2) and make enforceable contracts [12]. However, the idea that software systems are designed to satisfy requirements axiom is logically problematic in several ways.

First, if a requirement is something necessary for success, stating a set of requirements implies that we know what success means. In real projects, however, system goals are often unknown, ambiguous or disagreed on by stakeholders [13]. Rather, developers are faced with a situation that is perceived as problematic by different stakeholders in different ways [14]. If *success* is not understood or agreed, stating conditions necessary to produce the unknown success state is difficult if not impossible.

Second, even if success in the target domain is well understood, requirements elicitation may not produce requirements. Most people cannot produce statements that are "correct", "unambiguous", "complete" and "consistent" as the IEEE-830 standard [8, p. 4] recommends, even with expert help [12]. In the best case, the user hypothesizes that a feature is a necessary condition for success. However, the user may be incorrect in at least three ways – 1) the system could succeed without the feature; 2) the feature is inferior to an alternative not considered by the stakeholder; 3) the feature actually has a negative effect on system success. Perhaps more often, requirements elicitation identifies things users vaguely want or abstract goals. For example, in a study of a virtual learning environment (VLE) implementation, some users requested running two VLEs simultaneously despite multiple VLEs being identified as *the primary problem* with the status quo [13].

Third, designing involves manipulating elements whose interactions are not completely understood [15] or predictable [16]. In any realistically complex system, it may be impossible to tell whether any "requirement" is actually needed to achieve a particular goal or even whether it will have a positive effect. This is especially problematic as the word 'requirement' and the IEEE standard "the systems shall" phrasing both connote certainty.

Fourth, even if we clearly understand exactly what a system needs to accomplish, there may still be no requirements. Requirements are properties common to every system that will achieve a set of goals. If a goal can be accomplished by two systems, one with password-based identification and the other with biometric identification, then neither form of identification is a requirement, because there exists a potential solution with a different form. In situations when several heterogeneous solutions are evident, there may be too little in common to form a useful set of requirements. Moreover, supposing the developer has identified three workable solutions, in many contexts there is no way to know whether an unimagined fourth solution is still out there – a solution so different from current solutions as to invalidate any meaningful requirements (cf. [17]).

Furthermore, the idea that software systems are designed to satisfy requirements is contrary to many empirical observations [13]. Empirical studies of design professionals in many fields reveal that, in practice, design decisions are not driven by requirements specifications [18]-[20]. Rather, designers habitually treat problems as ill-defined, explore problems using solution conjectures and adjust goals and constraints as they go [18]. Research on creativity in requirements engineering (cf. [21]) is actually demonstrating that most "requirements" are actually design decisions.

However, the belief that software development is about satisfying requirements is intuitively appealing in several ways. First, it allows developers to offload much of the hard work of determining what to build – if the client provides faulty requirements, *that's their problem*. Second, it facilitates the logic of outsourcing – if no requirements are evident what would be the basis for outsourcing or tendering contracts? Third, a requirements specification conceals the innate tensions, disagreements and paradoxes within the problematic context, which eases developers' cognitive loads. Consequently, many practitioners and academics believe so strongly that requirements are central to SE that faced with the serious conceptual and empirical challenges identified above, they simply argue that meeting explicit requirements is the very definition of success. This oversimplified view of success has been thoroughly rejected in the project management literature, where success is now understood to include not only meeting schedule, budget and scope goals but also impacts on diverse stakeholders, market performance and learning [22]-[25].

Dogmatic adherence to the idea that design is driven by requirements is dangerous. First, it supports deleterious forms of tendering, outsourcing and contracting based on illusory requirements [12], [17]. When the outsourcee realizes that the specified system is not what the client really needs, they have three options: A) deliver an ineffective system that satisfies the contract but disappoints the client; B) deliver a better system, possibly at greater cost, and risk contractual penalties or a breach of contract lawsuit; or C) spend considerable time, resources and political capital trying to renegotiate the contract. Second, overconfidence in epistemically uncertain requirements likely interferes with developer creativity via myriad cognitive phenomena including myside bias, system justification, confirmation bias, default bias and design fixation [26]-[30]. Put simply, as requirements narrow the solution space, failing to question bad requirements may eliminate good designs. Finally, appreciation of the tensions, paradoxes and disagreements that are concealed by the requirements specification may be crucial for devising a good system.

## 2.2. Royce's Ontology is a Good Way of Conceptualizing Development Activity

Sim and Duffy [31] identified numerous generic engineering activities including "abstracting", "associating", "defining", "selecting", "searching", "scheduling" and "prioritizing." To reduce cognitive load and facilitate description, SE academics and practitioners organize these activities into higher level categories. For example, Table 1 shows overlapping categories between a curriculum guideline (SE2004 [4]), a process model (Waterfall [10]) a body of knowledge document (SWEBOK [3]), a system development method (RUP [32]) and Wikipedia's popular *Software Development Process Template* sidebar [33]. While each attempt produces minor variations and some are more comprehensive than others, categories equivalent to *requirements*, *analysis*, *design*, *coding and testing* consistently emerge. Although, the linear progression between phases dedicated to each category (as in Royce's Waterfall Model) is now widely recognized as idealistic, the categories themselves constitute a kind of implicit process theory, which we call *Royce's Ontology*. SE academics and practitioners widely take these categories for granted, i.e., they assume that Royce's Ontology is a good way of conceptualizing software development activity.

In Royce's Ontology, analysis refers to "detailed examination or study of something so as to determine its nature, structure, or essential features" and "breaking up of a complex whole into its basic elements or constituent parts" [34]. Design refers to specifying an object that is intended to accomplish goals in an environment, subject to constraints, where specifying could mean planning, prototyping or building the design object [35]. Coding refers to producing software programs and functions. Testing refers to evaluating properties (e.g. reliability, efficiency) of a software system (including its components and auxiliary artifacts) at different levels of analysis (e.g. unit, system, business).

Royce's Ontology is intuitively appealing in many ways. It is relatively simple and easy to teach. It employs everyday language, which is consistent with much of the SE literature, especially the methods literature (§2.4). However, it is also logically problematic and at odds with empirical evidence.

*Requirements* is a core activity in Royce's Ontology, yet, as discussed above, many projects have insufficient requirements to drive design. The remaining elements of *analysis*, plus *design*, *coding* and *testing*, in contrast, are not so fundamentally flawed; rather, it is conceptually decoupling them that is inconsistent with both empirical observation and recommended practice.

Extensive empirical research shows that conceptually decoupling analysis and design is problematic. Empirical studies demonstrate that professional designers and developers do "not keep means and ends separate"; rather, problem framing and problem solving are simultaneous and inextricably intertwined [20, p. 69]. Designers explore the problem space by generating design candidates and building software artifacts [18]. Developers' understanding of the problematic context and space of potential solutions co-evolve [18], [19], [36], [37]. Designers use solution conjectures better to understand problems [20].

Separating analysis, design and coding from testing is also illogical as testing may involve analysis, design and coding. Obviously, someone has to design the tests. With unit tests, for example, someone has to analyze the situation and design and code the tests. Furthermore, when a developer runs a code segment through a unit test suite and receives an error, sometimes the error is in the code, but other times it is in the test. Therefore, testing becomes entangled with coding and developers correct the test suite and the code through a mutual debugging process.

Furthermore, different kinds of testing are interconnected with design and coding. During coding, developers may employ myriad testing practices and tools including peer code reviews, static analysis tools and a debugger. Designers, especially visual designers, may employ user acceptance testing on lightweight prototypes and mockups to elicit user needs and preferences.

Similarly, conceptually decoupling design and coding implies that practitioners design systems without building them. Civil engineers and architects, among others, do this regularly. They can create sophisticated, mathematical models of contemporary structures accurately predicting how these structures will react to various loads, wind conditions and earthquakes. They can be reasonably confident of how their designs will behave prior to building them. Software engineers, in contrast, often develop complex systems – systems that exhibit properties not obvious from their constituent parts [16], [38]. As complex systems exhibit unpredictable behaviors, the only way to explore the design space is to build prototypes [13], [39]-[41].

Two interpretations of these logical and empirical challenges to Royce's Ontology are evident – one superficial and one deep. The superficial and unsurprising interpretation is that analysis, design, etc. are not mutually exclusive activities.

The less obvious, more fundamental conclusion is that Royce's Ontology is a dangerous dogma because its *vernacular* (not just the implied linearity of Waterfall) is defective and misleading. In Royce's Ontology, analysis etc. are categories of activities. "However, categories are useful only if they make it possible to infer further information, and only if they do so consistently" [42, p. 1040]. As Royce's categories are so overlapping (e.g. analysis and design) and contain such disparate activities (e.g. debugging vs. acceptance testing), we cannot draw meaningful inferences. For instance, if we know that "John is testing", we cannot infer whether John is an analyst, tester or programmer, whether he is working with a user, source code, a low-fidelity prototype, a debugger or a client, or whether this activity is automated or manual. Royce's Ontology is dangerous in several ways. For practitioners, conceptually separating analysis, design, etc. suggests that they can also be divided temporally, geographically or individually; e.g., we can do the analysis this month and start design next month, we can do the requirements in our UK office and coding in our Canadian office, our main team can develop the system while our quality assurance specialists handle the testing. While experienced practitioners may know such divisions are tenuous, remembering that the testers do some of the analysis and the analysts handle some of the testing requires a kind of perpetual doublethink. Meanwhile, the whole point of aggregating basic activities including "abstracting", "associating", "defining" and "selecting" into higher level categories including analysis and testing was to *reduce cognitive load*.

**Table 1: Influence of Royce's Ontology**

| Waterfall | SE2004 | SWEBOK | RUP Disciplines | Wikipedia |
|---|---|---|---|---|
| | | SE Economics | | |
| | | Software Configuration Management | | |
| | Computing Essentials | Computing Foundations | | |
| | Mathematical and Engineering Fundamentals | Mathematical Foundations | | |
| | Professional Practice | SE Professional Practice | | |
| System Requirements | | | Business Modelling | Requirements |
| Software Requirements | Software Modeling and Analysis | Software Requirements | Requirements | |
| Analysis | | | | |
| Program Design | Software Design | Software Design | Analysis and Design | Architecture |
| | | | | Design |
| Coding | | Software Construction | Implementation | Construction |
| Testing | Verification and Validation | Software Testing | Test | Testing |
| | Software Quality | | | Debugging |
| Operations | Software Evolution | Software Maintenance | Deployment | Deployment |
| | | | | Maintenance |
| | Software Process | SE Models and Methods | | |
| | Software Management | SE Management | | |

For researchers, poor classification of SE activities obscures the many fundamental SE phenomena that do not neatly map onto Royce's Ontology. For example, *coevolution* is a fundamental SE phenomenon where an engineer's attention oscillates between aspects of a problematic context and a space of possible solutions, gradually improving his or her understanding of each [18], [19], [37]. Despite coevolution being fundamental to thinking in the design discipline [18], [43], [44] it is barely mentioned in the SE methods literature or model curricula [6]. Consequently, Royce's Ontology inhibits understanding and application of alternative theories of software development, including Sensemaking-Coevolution-Implementation Theory [13], [36], [45], [46], which attempts to organize SE activities into more cohesive, less overlapping categories.

### 2.3. Software Engineers Design Software

The official IEEE definition of *software engineering* reads in part, "the application of a systematic, disciplined, quantifiable approach to the development, operation, and maintenance of software" [47]. The assertion that SE practitioners design and build *software* systems appears so uncontroversial that it is rarely questioned. However, asserting that SE is only about software implies that software design is fundamentally different and separate from hardware design and social system design. We even invented the term *virtual* to distinguish software artifacts from their material counterparts, i.e., the hardware systems used by the software and the social systems that use the software. This *virtual-material dualism* comprises two basic assumptions:
1. Software systems are fundamentally different from the hardware systems they run on
2. Application software is fundamentally different from the social systems that use it

Virtual-material dualism is appealing in many ways. From the perspective of an application developer, both hardware and social systems are often predetermined or relatively unchangeable. Conceptual boundaries drawn around the virtual software coincide with conceptual boundaries between what is and is not under the developer's control. Imagining these as non-permeable boundaries helps developers manage project complexity, scope creep and cognitive load. Similarly, from the perspective of an application user, events occurring on-screen do not appear as real as events in the physical world. In video games, for instance, many of us perform actions *for fun* that would be clinically traumatic in the real world. Moreover, many users readily distinguish between hardware and application software errors; e.g., when a tablet application gives an error, we first contact the people who made the application, not the people who made the tablet. However, virtual-material dualism is logically problematic in several ways.

First, by designing application and enterprise software, developers often implicitly design parts of the social system where the software will be implemented. For example, at the time of writing, the plagiarism detection system Turnitin allows teachers to grade papers while easily switching between original and originality-analysis views. However, the student's name is visible in all views, preventing anonymous grading. In this way, Turnitin's design exerts influence on the broader social system. Of course, the user can work around Turnitin's design by downloading and anonymizing the papers; however, Turnitin's designers have still implied an intended design dimension of the social system. In this way, the design artifact boundary may not align with the virtual-material boundary. Similarly, when a company adopts enterprise resource planning software, it is often easier to adjust business process to the software than to adjust the software to existing business processes [48]. This dynamic gives software developers substantial control over business processes.

Second, in some contexts hardware and software design are tightly connected. For example, to develop the Chromebook, Google simultaneously developed an operating system (linux-based ChromeOS) and a reference hardware design (the CR-48 prototype). Clearly the hardware design influenced the software design, e.g., USB ports necessitate software drivers. However, the software design also influenced the hardware design, e.g., due to the web-centric nature of ChromeOS, the CR-48 keyboard replaced the *caps lock* button with a dedicated *search* button. Other examples include many embedded systems (e.g. traffic lights) where hardware-software co-design is common [49]. Here again the boundary between the design artifact and its context does not align with the virtual-material boundary.

Third, the system boundaries perceived by users may not align with the virtual-material boundary. For example, when a non-technical user cannot connect to WiFi, he or she does not readily distinguish between the modem, modem firmware, router, router firmware, network protocols, local wireless hardware, wireless drivers and the browser. The non-technical user lumps them all together into a single system by saying 'the Internet is broken'. The system boundary is perceived between "the Internet" and the user, not between the hardware and the software. Similarly, pervasive games often combine material and virtual activity [50]. The boundary between the game world and the "real" world is therefore not aligned with the virtual/material boundary [51].

Neither the idea of hardware-software co-design nor the idea of the design artifact being a sociotechnical system rather than purely software are new [14], [49], [52]. However, outside of particular subfields (e.g. embedded systems, information systems) both SE practitioners and academics often appear to act as though software development is fundamentally disconnected from hardware and social systems. Some have even argued that modern software complexity is rooted in hardware designer's propensity to leave all but the most basic operations to software [53]. Moreover, diverse research on designing human activity systems (e.g. [14], [54]) including work systems (e.g. [52], [55]) and service systems (e.g. [56]) has had limited impact on standard curricula or bodies of knowledge (cf. [3], [4]). Industrial design and business process design, for example, appear rare in SE educational, theoretical and practical discourse. Even increasing interest in user experience design does not go as far as rejecting virtual-material duality and recognize that designing the work software may entail designing the work system that employs it [57]. Yet designing the work system through the software is exactly what SAP and other enterprise resource planning software appears to do.

The virtual-material duality dogma is dangerous in at least two ways. First, failing to consider the complex interaction between a human activity system and the software systems that support it can have serious consequences. For example, the software-based Integrated Children's System was developed to support social workers in the UK. "By attempting to micro-manage work through a rigid performance management regime, and a centrally prescribed practice model", the Integrated Children's System "has disrupted the professional task, engendering a range of unsafe practices" [58, p. 405]. The software effectively created a toxic environment that increased the probability of tragedies including the death of the 17-month-old British boy known as Baby P [59]. In this instance, the software not only was ineffective but also actively damaged the design of the human activity system.

Second, conflating *software* and *design artifact* may unintentionally and unnecessarily narrow the solution space, handicapping innovation. Suppose a software developer distinguishes between three kinds of systems – hardware, software and social – and equates the software system with the design artifact. When designing, however, the important boundary is between design artifact and context; i.e., what the designer can and cannot manipulate. This may lead developers to create complex, expensive software solutions where simple, economical hardware or social changes would suffice. Similarly for researchers, defining the design artifact as software independently of a particular context may obscure important social dimensions of software adoption and implementation, e.g., enterprise system implementation failing because the organization did not make the needed business process changes rather than because of particular features of the software.

The virtual-material duality dogma may have even more subtle and insidious consequences. Default bias is a common psychological tendency to select options which are presented as the default or status quo [26], [60]. The combination of time constraints, default bias and simple laziness gives software developers often unrecognized power to influence user behavior. Rather than explicitly designing human activity system, software developers can implicitly manipulate human behavior by making preferred actions easier or framing them as defaults. For instance, if Turnitin made anonymous grading the default option, many more academics would likely take that route. This presents two interconnected dangers – 1) negative unintended consequences of design decisions, 2) intentional manipulation of users. Of course, these are difficult to distinguish from the outside – we do not know whether Turnitin's developers are implementing a principled stance against anonymous marking or if it is simply an unintended consequence.

## 2.4. Methods are Used, Effective and Generalizable

The concept of a system development method(ology), or simply a *method*, occupies a deeply privileged position in the academic discourse [61]. A method (e.g. Scrum, Lean, Crystal, Kanban, Rational Unified Process, PRINCE2) is a collection of interconnected, complex, abstract prescriptions concerning how to build a system effectively and connotes "an orderly, predictable and universal approach" [61, p. 54]. The concept of method thus entails three assumptions – 1) methods are used, 2) using a method positively affects SE success, 3) each method generalizes to a range of contexts (not that one method can be effective in every conceivable project, just that a method applies to many projects).

Although common in the literature (cf. [3], [4], [61]-[67]), these assumptions are so rudimentary that they are rarely stated explicitly. However, they are clearly fundamental assumptions of method engineering (e.g [68]), software process improvement (e.g. [69] )and process maturity (e.g. [70]).

The assumptions that methods are used, effective and generalizable are intuitively appealing in several ways. Methods can be thought of as being similar to tools: a debugger, for instance, is clearly used, often effective and applicable to myriad software projects. Methods can be thought of as being similar to algorithms: collaborative filtering, for example, is used by recommender systems, is fairly effective and applies to a wide range of item classes. Finally methods can be thought of as being similar to highly structured procedures in other domains; for instance, a banker may follow a precise procedure to determine whether a client is eligible for a particular mortgage.

However, conceptualizing methods as similar to tools, algorithms and procedures is logically problematic, since methods are fundamentally different. A method is a *collection* of interconnected, *complex, abstract* prescriptions. As methods are collections, developers may choose to follow *some* of the prescriptions *some* of the time. As prescriptions are complex, developers may partially comply; e.g., peer programming is not simply having two developers share a workstation [11]. As prescriptions are abstract, they may be instantiated differently in different contexts; e.g., working out how to conduct a daily stand-up meeting (Scrum) in a geographically distributed team.

Unsurprisingly, then, substantial empirical research demonstrates that methods are neither effectively nor extensively used [71]-[76], rarely used as intended [73]-[78] and almost never empirically validated [62]. Developers may claim to follow a method or even fake adherence while practically ignoring it [65], [72], [79]}. Similar methods in similar situations may produce opposite results [80] while specific methods may be unsuitable for certain individuals [81]. Organizations may change so quickly that any long-term method becomes ineffective [67], [82]. Rather than *use* methods, practitioners create "methods-in-use" [83], which significantly differ from methods(-in-books). Moreover, practitioners often deviate from methods not from ignorance but from recognition of the method's misalignment with their context [75], [78]. This has led many to the conclusion that *methods are fictions* (e.g. [61], [79], [84]). Yet the *methods are used, effective and generalizable* dogma remains despite being logically problematic and contradicted by substantial empirical evidence.

Method-fictions may be useful for practitioners in that trying to apply them produces more rational design processes [79]. However, method-fictions, and this dogma, are also dangerous. First, many method-fictions foster an over-rationalized worldview that obscures fundamental SE phenomena including improvisation [13], [85]-[87], amethodical development [61], [67] and goal disagreement. Second, method-fictions contribute to a non-empirical culture within SE. When mechanical engineering researchers design a new piston, they are expected to provide empirical evidence that the new piston outperforms existing pistons, but when SE researchers develop a new method, no one expects them to empirically demonstrate the new method's superiority to existing methods. This has led to a proliferation of thousands of methods [64], [88] with little empirical knowledge of their relative efficacy [62]. Third, the privileged status of method in the literature leads researchers to theorize on the basis of method-fictions rather than empirical realities. For example, Sjøberg et al. present a theory of "UML-based development" which posits that "the use of a UML-based method increases costs [and] … positively affects communication, … design, … documentation … [and] testability" [89, p. 323]. This formulation glosses over the empirical reality that, in practice, UML is used rarely, selectively and informally [74], [75]. While their hypotheses are not necessarily wrong, they are based on an oversimplified notion of use, which obscures the diversity of interactions between UML concepts and real-world development.

Finally, methods, or method-fictions, constitute an accumulation of other dogmas, such as those we have discussed in this section; many of the methods prescribe that software systems should be designed to satisfy pre-defined system requirements, utilize Royce's Ontology and encourage a virtual-material dualism. They also serve to disseminate such dogmas further as SE practitioners and researchers move to adopt the methods, thus encouraging a persistent belief in SE dogmas.

## 3. EVIDENCE-BASED SOFTWARE ENGINEERING

The previous section discussed some core assumptions in SE that have endured despite being logically problematic and contrary to many empirical observations. One factor that may substantially contribute to the resilience of these dogmas is the gap between SE research and practice, which continues to widen due to "a tremendous lack of empirical evidence regarding the benefits and limitations of new SE methods and tools on both sides" [90, p. 13]. Currently SE researchers may design tools and methods without knowing that they do not work well when adopted by practitioners, and practitioners may adopt tools and methods without knowing that they did not work well for other adopters. Widely-held dogmas provide a sense of security that "everyone does it like this". This section therefore discusses one approach to close the research-practice gap and rejuvenate empiricism on both sides – evidence-based practice (EBP).

### 3.4. Evidence-Based Practice

EBP, in its simplest sense, refers to a form of professional activity where practical decisions are based on empirical research. In reality however, EBP is a relationship between communities of researchers and practitioners comprising roughly three parts – 1) empirical research on practically relevant problems, 2) systematic literature reviews (SLRs, below) written for non-academics, 3) professional activity guided by SLR conclusions. Successful EBP therefore requires two forms of 'readiness' [91] – researchers ready to study areas of concern to practitioners, and practitioners ready to integrate scientific evidence into their practice.

An SLR analyzes and synthesizes published empirical studies concerning a method, tool or technology to establish the evidence for its effectiveness. The search strategy for discovering primary empirical research is made traceable and repeatable (e.g. all search terms and databases used are listed), the criteria for deciding which studies are of high-enough quality to be included in the synthesis are defined and applied, and the findings of the high-quality empirical studies are synthesized so that conclusions can be drawn about the effectiveness in-use of the artifact under consideration [92]. SLR summaries are often written for practitioners in non-academic language.

EBP originated in medicine. For improving practitioners' decision-making, which had previously been based on habit, prejudice or imperfect knowledge of relevant research [93], evidenced-based medicine is considered one of the most important medical innovations since 1840 [94]. As doctors often lack the time or skills to process multitudinous papers largely written *by* academics *for* academics, EBP forms the aforementioned three-stage process. First, medical researchers conduct and publish basic and applied studies in formats largely intended for other researchers. Second, researchers conduct *SLRs* and publish non-academic summaries in a free, open, international, online knowledgebase (the Cochrane Collaboration; www.cochrane.org). Finally, doctors combine guidance from the knowledgebase with the values and preferences of patients to determine the best treatment [93].

Of course evidence-based practice is not without controversy. For example, it is dependent on the quality of empirical evidence produced, and it may be criticized for bias toward experimental evidence and sensitivity to publication bias [95]. While Cochrane reviews, for example, are generally better than industry-supported reviews [96], some Cochrane reviews still overstate the empirical support for treatments [97]. EBP is also criticized for preventing physicians from tailoring treatments for specific cases; however, this reflects a common misunderstanding of EBP – EBP creates guidelines, not rules, which practitioners are expected to adjust for unique circumstances (Figure 1).

Despite some criticism, EBP has expanded beyond purely clinical treatments to healthcare management and policy; e.g., strategies to change organizational culture to improve healthcare performance [98], and approaches for promoting information technology adoption by healthcare professionals [99]. EBP has also spread to other disciplines, such as social policy [100], librarianship [101] and education [102]. The Campbell Collaboration (www.campbellcollaboration.org) now houses SLRs in education, social welfare and crime and justice.

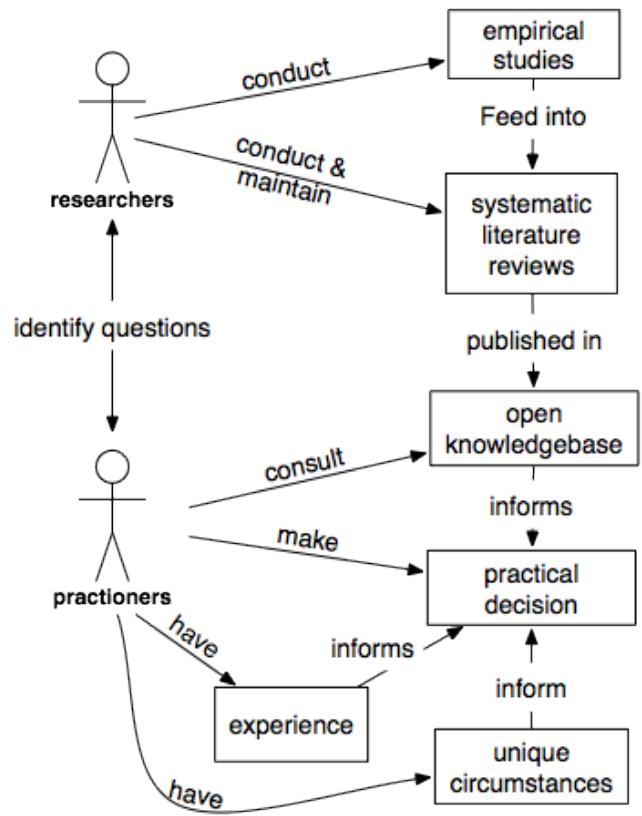

**Figure 1: Overview of Evidenced Based Practice**

### 3.5. Evidenced Based Software Engineering

An evidence-based software engineering community (EBSE) has also developed, which is exploring how EBP can be adapted for SE [103]. Via a series of conference workshops and publications this community has promoted EBSE and published SLRs for some SE beliefs and practices (e.g. [104], [105]). A web repository for EBSE has also been established (www.ebse.org.uk).

However, EBSE adoption remains slow and limited to a small number of enthusiasts. A 2011 study found only 49 SLRs in SE since the beginning of 2005, and only half of those contained a synthesis of the empirical evidence – the remainder simply identified the empirical studies available on a given topic [106]. The EBSE repository has apparently not been updated since August 2012. Its SLRs have not been summarized for practitioners; instead the citation details and abstract of the SLR-based paper written for academics is provided. Sometimes a one or two sentence summary of the review's findings is included, but this is clearly insufficient to support practitioner decision-making.

A more pronounced movement towards EBSE would facilitate differentiating between unempirical dogma and veracious theory. Theoretical principles facilitate academic consensus building and protect practitioners from con men dressed as "Agile coaches" and "software process improvement consultants". An EBSE culture would be a powerful inoculation against the unempirical fads that pervade SE [107]. It would also increase pressure to base educational curricula on evidence rather than custom and opinion. Moreover, EBSE necessarily addresses the research-practice knowledge gap as practitioners are familiar with more research and researchers face more practical questions. More fundamentally, without SLRs we cannot definitively answer even simple questions and this impotence breeds dangerous dogmas including those above.

## 3.6. Recommendations for Progressing EBSE

Transitioning to an EBP paradigm, however, entails myriad nontrivial social, political and methodological challenges. To move to EBP, researchers need to address practical problems and provide more and better empirical evaluations of proposed technologies and practices. This means that SE students, researchers and reviewers need more and better training in the full range of empirical methods appropriate to SE, from surveys and experiment to ethnography and action research. Journals and conferences must also demand such evidence. The ICSE 2014 call for submissions invites methodological papers that "provide convincing arguments, with commensurate experiences, why a new method is needed and what the benefits of the proposed method are." Arguments and experiences alone are not good enough in engineering, management, medicine or psychology – we see no good reason for SE to accept a lower standard. For this to work, reviewers will also need to focus more on evidence and methodology (and less on literature reviews, style and framing). Promotion and tenure committees and research funding councils should also acknowledge that building and evaluating takes longer and represents a greater contribution than building alone.

More empirical research supports more secondary analysis. However, as performing SLRs is methodologically difficult, SE researchers need greater training in meta-analysis and thematic synthesis. New methods will also be needed, to synthesize the heterogeneous data commonly found in empirical SE studies that do not originate from randomized controlled trials as used in clinical medicine [108], [109]. As conducting and maintaining SLRs is time-consuming, promotion and tenure committees and research councils must recognize SLRs as major scientific contributions.

SLRs must be summarized in a format suitable for practitioners, and should be made freely available via a website, such as the EBSE website that has already been initiated. Practitioners need exposure to the SLR repository and its benefits, and training in how to effectively combine guidelines from SLRs with experience and circumstance to act effectively. SE educational programs should make the SLR repository the go-to resource for SE students. Industry standards and educational curricula should reflect not only evidence from SLRs (rather than popular opinion and dogmas) but also how to apply SLR guidance in practice.

As the number of SLRs increases practitioners could increasingly base their decisions on the available evidence, combined with their knowledge of their own particular context. This will alter SE's implicit model of scientific communication: the current reports *by* researchers *for* researchers will also be mediated via summaries of systematic reviews into an accessible format for practitioners, which will bring research and practice much closer together. The gap between them could also be bridged if researchers more frequently conducted engaged research, i.e., empirical studies in practitioner organizations.

Highly effective EBSE entails even more difficult challenges. For example, comparing heterogeneous technologies and practices requires common agreement on a top-level dependent variable. Intermediate variables including developer productivity, lines of code and number of bugs lack the unambiguous finality of morbidity and mortality. We therefore need a better understanding of what constitutes "software engineering success" [25], [110]. Furthermore, the entire academic publishing apparatus fundamentally undermines SLR accuracy through publication bias [111], leading some to argue for revolutionary changes including simply publishing all empirical research [95].

This agenda offers the prospect of a SE discipline which requires empirical evidence for the efficacy of new technologies and practices *before* their adoption and standardization, and emphasizes cumulative knowledge development rather than doggedly clinging to dogmas.

## 4. CONCLUSIONS

This paper makes two interconnected contributions. First, it presents four dogmas, i.e., widespread non-empirical beliefs, which pervade the SE discourse despite being logically problematic. Our deconstruction of these dogmas suggests the following.

1. In many SE projects the "requirements" are not simply flawed but entirely fictitious, and these fictitious requirements mislead and undermine good design.

2. Conceptually organizing SE activity into categories including *analysis*, *design*, *coding* and *testing* is confusing because these categories are not mutually exclusive and do not support inferences about their component activities.

3. Conceptually demarcating *the software* from hardware and human activity systems may lead to unnecessarily complex software, reduced innovation and unintended consequences.

4. System development methods are fictions that impair research and practice by over-rationalizing, oversimplifying and obscuring paradoxes, conflicts and improvisation.

Second, it suggests that the endurance of these dogmas results from the research-practice gap, which may be addressed through evidenced-based practice. We furthermore suggest an agenda for transitioning to EBP, which necessitates substantial changes including better empirical training for academics, more academic credit for writing systematic reviews and greater focus on evidence in peer-reviewing, publication, standards, curricula and bodies of knowledge.

In this paper we have pointed to empirical evidence that contradicts four common SE assumptions, to raise awareness of dangerous dogmas. We recognize that there may also be some empirical evidence to support these core assumptions. Our paper will have succeeded if it stimulates researchers to examine the empirical evidence for these assumptions, both for and against, and reach a conclusion about their validity; in other words, if researchers are inspired to carry out a SLR and contribute to evidence-based practice. Further research should also investigate other core assumptions and non-empirical concepts in SE.

Some feel that empiricism has, in principle, a quite limited role in SE research (e.g. [112]). If SE wants to be, and to be regarded as,

a scientific discipline, such positions bode ill indeed. Empiricism is one of the fundamental characteristics that distinguishes science from pseudoscience [113], [114]. Without embracing empiricism, SE will likely continue to rely on dogma instead of well-founded theories that are supported by evidence. Without fostering evidenced-based practice, SE will likely continue to flounder in the gap between what ought to work and what actually works. Continuing to foster non-empirical concepts and weakly validated technologies and practices may lead to widespread belief that most SE is pseudoscience and eventually to marginalization of the entire community.